\documentclass[journal]{IEEEtran}

\usepackage{cite,algorithm,algorithmic,psfrag,bm,url}
\usepackage[dvips]{graphicx}
\usepackage[cmex10]{amsmath}

\ifCLASSINFOpdf
\else
\fi

\begin{document}
%
\title{Black-box Modeling and Compensation of Bursty Communication Signals in RF Power Amplifiers with Power-Dependent Parameters}
%
%
%

\author{Ali~Soltani~Tehrani,
        Haiying~Cao,
        Thomas~Eriksson,
        and~Christian~Fager
\thanks{A. Soltani Tehrani and T. Eriksson are with the Department
of Signals and Systems, Chalmers University of Technology, Gothenburg,
Sweden e-mail: ali.soltani@me.com, thomase@chalmers.se.}
\thanks{C. Fager is with the Department of Microsystems and Nanotechnology, Chalmers University of Technology, Gothenburg,
Sweden e-mail: christian.fager@chalmers.se.}
\thanks{H. Cao is with Ericsson AB, 164 80, Stockholm,
Sweden e-mail: haiying.cao@ericsson.com.}
}

\maketitle

\begin{abstract}
This paper presents a new black-box technique for modeling long term memory effects in radio frequency power amplifiers. The proposed technique extends commonly used behavioral models by utilizing parameters that dynamically change depending on a long term memory effect while keeping the original model structure intact. This enables us to accurately track and model transient changes in power amplifier characteristics that vary slowly and are induced by the input signal.

Identification of long term memory effects is discussed and an iterative identification algorithm for the model parameters is proposed. The model is experimentally tested on a 100 Watt Doherty power amplifier with a 4 MHz Gaussian noise signal that has a step--like change in the amplitude, representative of a realistic communication signal with bursty behavior and a 20 MHz 3GPP LTE test data. Results of behavioral modeling show a 2-2.5 dB and 5-6 dB improvement in average and peak NMSE modeling performance respectively, which shows the suitability of the technique to model bursty signals.
\end{abstract}


\IEEEpeerreviewmaketitle

\section{Introduction}


%
As modern communication systems became more spectrally efficient by utilizing amplitude--varying signals, the distortion caused by power amplifier (PA) memory effects on the signal has increased. Deriving and comparing behavioral models that can model these effects is a pre-requisite to be able to perform system level simulations or to be able to successfully linearize PAs in wireless transmitters, which has therefore been focus of much research \cite{pedro, soltani, isaksson}.

Researchers have shown that there are mainly two categories of memory effects that degrade the communication signal: short-term memory effects which are normally attributed to electrical memory effects generated from the RF behavior of the matching networks, and long-term memory effects which are due to biasing circuits, temperature drifts and trapping effects \cite{vuolevi,ngoya,ku2003}. In \cite{thermal_boumaiza}, it was shown that for communication signals with wide bandwidth and relatively constant average power, the electrical short term memory dominated the behavioral modeling performance, and the focus in the literature has thus been on modeling short term memory effects. However, in \cite{bosch} it was shown that long term memory effects cause the most critical signal distortion, as they are harder to deal with using common predistortion algorithms.

As mobile usage patterns shift from simple voice calls to data transmission and packet--based systems \cite{coda}, communication signals have moved from relatively constant power signals to signals in bursts and packets in some standard scenarios. These bursty signals induce longer memory effects in power amplifiers by, for example, changing the instantaneous temperature in the PA. Traditionally, behavioral models have been developed to assume relative constant power (and temperature for example), while leaving parameter adaptation techniques to update for changes in temperature. In order to account for long term memory effects in the literature, different approaches have been taken. In 
\cite{ku2003}, pruning behavioral models to include sparse delay has been an approach to model long term memory effects. However, only utilizing sparse taps may not be sufficient to track changes in power amplifier behavior, and better techniques need to be developed. Thermal networks are utilized in \cite{thermal_boumaiza,thermal_mazeau} to model and compensate for thermal gain variations. In \cite{crespo2010}, a  behavioral model to include long term memory effects is developed using gray-box knowledge of the thermal filter of the PA and a lower complex version of this model is presented in \cite{crespo2011}. In \cite{ngoya}, a different approach to is proposed, by utilizing continuous time models. This model requires a relatively complicated identification procedure with numerical techniques and suggests using cubic-spline functions for the long term memory component.

In this work, which extends \cite{soltani2012}, we propose a novel approach where the parameters of the behavioral model are assumed to be 
dependent on a state parameter like average power, and use it to track dynamic changes of the PA behavior when fed with bursty data. 
This enables us to keep the model structure of commonly used models in the literature (like the Volterra model, memory polynomial model, etc) intact and successfully improve their performance with regards to these type of memory effects in the PA. The proposed modeling technique works can effectively be seen as a pro-active parameter adaptation scheme, that can track changes in average power level,
and may therefore relax the need for faster adaptation in DPD applications. Further, the proposed modeling technique is easily compatible with traditional adaptation algorithms, and can be incorporated jointly.

In this paper an iterative identification technique is presented for parameter estimation. The model is evaluated with a modulated signal with a sudden change in power level, which represents the bursty data of new mobile usage trends, and the performance improvement is analyzed. Further model generalization is done and tracking multiple long term memory effects is also presented.

The paper is organized as follows, in Section II the model formulation is presented and the model complexity is analyzed. In Section III, an iterative procedure of identification of such models is presented. In Section IV the measurement setup is introduced and the results of modeling and predistortion using the proposed model are presented and analyzed. Finally, in Section V, the results are discussed and conclusions are drawn.

\section{Modeling Bursty Communication Signals}
Bursty communication signals, due to the change in average power levels, commonly inflict changes in the PA behavior such as temperature drifts /cite{soltani2012}.
These effects are commonly very long term and will be referred to as such, and need to be modeled and compensated for. This section introduces the proposed behavioral modeling technique.

In the case of linear--in--parameters discrete--time behavioral modeling of power amplifiers, it is common to construct the model by generating nonlinear basis functions of the input and then identifying the kernels or parameters of the model. This can be written in simple matrix form as
\begin{align}
y[n] = \mathbf{H}_{x[n]}\boldsymbol{\theta},
\label{traditional}
\end{align}
where $y[n]$ is the baseband output sample of the model, $\mathbf{H}_{x[n]}$ is a vector of the basis functions consisting of different nonlinear and memory orders of the baseband input signal $x[n]$, and $\bm{\theta}$ is the model parameters vector. Many of the models proposed in the literature can be written as (\ref{traditional}), examples of which include Volterra and Volterra--based models, the memory polynomial model \cite{kim}, the generalized memory polynomial \cite{morgan}, and other similar models \cite{pedro}. In previous works, including long term memory effects have generally resulted in developing more complex $\mathbf{H_{x}}$ matrixes that remain linear in parameters \cite{crespo2010} or utilizing continuous-time models to integrate the long term memory effect in $\mathbf{H_{x}}$ \cite{ngoya}. In this section, the we propose a modeling technique that introduces the long--term memory effect in the parameters of the behavioral model $\bm{\theta}$ instead.

\subsection{Model formulation}
The black--box interpretation of utilizing time-dependent parameters for behavioral modeling is that we assume that the changes in power amplifier behavior -- such as biasing effects, temperature drifts, etc -- alter the parameters of the model while keeping the model structure unchanged. Effectively, this is the same assumption that is done in adaptive digital predistortion, where the model structure remains unchanged and only parameters are updated.

The proposed model structure can thus be written as
\begin{align}
y[n] = \mathbf{H}_{x[n]}\bm{\theta}\mathbf{(s)},
\label{ordinary}
\end{align}
where the model structure remains similar as before (it can be the Volterra model, a memory polynomial or any other commonly used behavioral models linear in parameters), and $\bm{\theta}\mathbf{(s)}$ are the parameters of the behavioral model that are now a function of the instantaneous state of the power amplifier, $\mathbf{s}$.

A simple first order approximation of the dependence of the parameters of the model on this state vector $\mathbf{s}$ can be made. This is equivalent to a first order Taylor expansion of the parameters $\bm{\theta}\mathbf{(s)}$ around the state vector $\mathbf{s}$ and the proposed model can be written as
\begin{align}
y_{\text{model}}[n] = \mathbf{H}_{x[n]}\left(\bm{\theta}^{(0)} + s[n]\bm{\theta}^{(1)}\right), \label{model_structure}
\end{align}
where $\bm{\theta}^{(0)}$ are the parameters of the behavioral model independent of the state, $\bm{\theta}^{(1)}$ are the dynamic parameters dependent on state $s[n]$, and $\mathbf{H}_{x[n]}$ is identical to (\ref{ordinary}) are the same columns of common behavioral model structures proposed in the literature.

Written in this general form, it can be noticed that models that are developed for modeling short term memory can be extended to include longer term memory with relative ease.

\subsection{Long term memory state}
In the literature it has been shown that long term effects of PAs can be modeled as filtered versions of the magnitude of the input signal $|x[n]|$ or the power $|x[n]|^2$ \cite{ngoya,crespo2011,ngoya2009}. We will use this assumption and setup the state parameter $s[n]$ of the PA as a filtered version of $|x[n]|^2$.

\begin{figure}
\centering
\psfrag{x}[c][c][1]{$x[n]$}\psfrag{y}[c][c][1]{$y[n]$}\psfrag{s}[c][c][1]{$s[n]$}\psfrag{A}[c][c][1]{$\mathbf{H}_x\bm{\theta}(s)$}
\psfrag{B}[c][c][1]{$G(\omega)$}\psfrag{C}[c][c][1]{$|\cdot|^2$}\psfrag{i}[c][c][0.75]{Input signal} \psfrag{o}[c][c][0.75]{Output signal} \psfrag{f}[c][c][0.75]{Low pass filter} \psfrag{p}[c][c][0.75]{Signal power} \psfrag{d}[c][c][0.75]{Long term}\psfrag{e}[c][c][0.75]{memory state}\psfrag{m}[c][c][0.75]{Model basis}
\includegraphics[width=0.8\columnwidth]{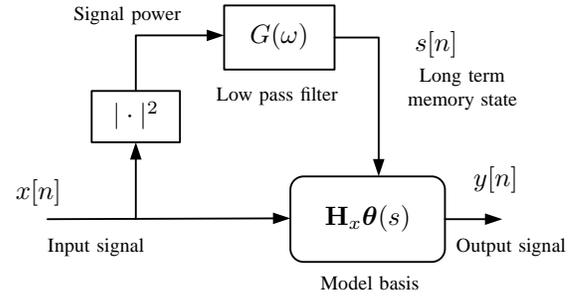}
\caption{Block diagram of the proposed modeling technique. $G(\omega)$ is a low--pass filter.} \label{setup1}
\end{figure}

Since $s[n]$ tracks slowly--varying long term effects, we model the long-term behavior with an autoregressive (AR) (\ref{AR}), autoregressive and moving average (ARMA) (\ref{ARMA}) and finite impulse response (FIR) from \cite{soltani2012} (\ref{FIR}) low--pass linear filter. The block diagram of the proposed modeling technique is shown Fig.~\ref{setup1}.

The AR filter will be written as AR($T$),
\begin{equation}
s_{\text{AR}}[n] = |x[n]|^2 + \sum_{k=1}^{T} \alpha_k s_{\text{AR}}[n-k], \label{AR}
\end{equation}
where $T$ is the number of poles and $\alpha_k$ are the poles in the AR model. The ARMA model can be written as ARMA($T$,$Z$),
\begin{equation}
s_{\text{ARMA}}[n] = |x[n]|^2 + \sum_{k=1}^{Z} \beta_k |x[n-k]|^2 + \sum_{k=1}^{T} \alpha_k s_{\text{ARMA}}[n-k], \label{ARMA}
\end{equation}
where $T$ is the number of poles, $Z$ is the number of zeros, and $\beta_k$ and $\alpha _k$ are the zeros and poles respectively. The FIR model from \cite{soltani2012} will also be used for comparisons, written as
\begin{equation}
s_{\text{FIR}}[n] = \frac{1}{N}\sum_{k=0}^{N-1} |x[n-k]|^2,
\label{FIR1}
\end{equation}
where $N$ is the window size of the moving average FIR filter.

In the case of AR(1) filters, the approximate effective memory length, $\tau$, can be calculated from the pole as
\begin{align}
\tau = \frac{1}{1-\alpha}
\end{align}
and for FIR filters the size of the window $N$ represents the long--term memory length.

\subsection{Model complexity}
In terms of running complexity \cite{soltani}, after choosing a behavioral model structure for the model basis in Fig.~\ref{setup1}, for the same nonlinear orders and memory depths, the proposed model has twice the number of parameters compared to the original model structure. The only other added calculation needed per coefficient is to find $s[n]$, which is one addition and one substraction per coefficient per pole/zero and one complex multiplication. It can be noticed that computing the power amplifier state adds very little to overall complexity to the behavioral model, as long as the number of poles/zeros are not high.

\section{Model Identification}
In this section, identification of the different model parameters is discussed. To better explain the procedure, a simple memory polynomial model \cite{kim} is used as the model structure $\mathbf{H}_{x[n]}$, and an ARMA(1,1) filter is used for modeling the low pass filter. Under these assumptions, the proposed model (\ref{model_structure}) can be expressed as
\begin{align}
y[n] = \sum_{p=1}^P\sum_{m=0}^M \left(\theta_{p,m}^{(0)} + s[n]\theta_{p,m}^{(1)}\right)x[n-m]\left|x[n-m]\right|^{p-1} \label{model_structure_mp}
\end{align}
where with $s[n]$ can be written as (\ref{ARMA})
\begin{align}
s[n] = |x[n]|^2 + \beta |x[n-1]|^2 + \alpha s[n-1]
\label{ARMA_mp}
\end{align}
and $\beta$ and $\alpha$, $\theta_{p,m}^{(0)}$ and $\theta_{p,m}^{(1)}$ are the unknown parameters.

From this formulation it should be noticed that while the output is linear with respect to the $\theta$ parameters, 
the parameters of the ARMA filter $(\alpha,\beta)$ (or the FIR filter parameter $N$ when using an FIR filter instead) 
appear nonlinearly in the output model and normal linear identification techniques cannot be used.

\subsection{Algorithm Outline}
As we have two sets of parameters to identify ($\bm{\theta}$ and $[\alpha,\beta]$) and only one set is linear with respect to the output ($\bm{\theta}$), iterative solutions are needed. In this work a simple recursive algorithm to iterate between solutions until all parameters converge.  The identification algorithm is shown in Algorithm \ref{algo}.
\begin{algorithm}
                \caption{Iterative Identification}
                \label{algorithm}
                \begin{algorithmic}
                                \STATE \emph{Step 1}: Find the initial values $[\alpha,\beta]_{\text{init}}$
                                \STATE \emph{Step 2}: $[\alpha,\beta]_{k} = [\alpha,\beta]_{\text{init}}$
                                \REPEAT
                                                \STATE \emph{Step 3}: Find $s[n] = \alpha_{k} s[n-1] + |x[n]|^2 + \beta_k|x[n-1]|^2 $
                                                \STATE \emph{Step 4}: Fix $s[n]$, identify $\mathbf{\theta^{(0)}}$ and $\mathbf{\theta^{(1)}}$ from least-squares estimation as shown in (\ref{thetaident})
                                                \REPEAT
                                                    \STATE \emph{Step 5}: Fix,$\bm{\theta}$, update $[\alpha_{k+1},\beta_{k+1}]$ from $[\alpha_{k},\beta_{k}]$ with Gauss-Newton steps, as shown in (\ref{a})
                                                \UNTIL{$[\alpha,\beta]$ converge}
                                \UNTIL{performance converges}
                \end{algorithmic}
                \label{algo}
\end{algorithm}

\subsection{Initialization}
Since an iterative identification method is used to identify the parameters sets, and the long--term memory filter 
parameters do not appear linearly in the output for the AR, ARMA and FIR filers usd, having proper initial values 
is important to improve the speed of convergence for identification. 

The first step is to initialize the parameters of the ARMA
filter. In this work, this is done with two--tone measurement in a procedure similar to \cite{ku2003}, but developed
in the framework of the proposed model. Once this initialization values are found, they can be used in the iterative
algorithm for identification. The two--tone setup and measurements are explained in the Appendix. 

Once the parameters are initialized, they can be used to calculate $s[n]$ using (\ref{AR}), (\ref{ARMA}) or (\ref{FIR1}). After 
the this state parameters is established, it can be used as for the next identification step.

\subsection{Model basis parameters}
Once the initial $s[n]$ is calculated from Step 3, it can be noticed that the parameters $\bm{\theta}$ are the only remaining unknowns. In order to identify these parameters, the modeling equation (\ref{model_structure_mp}) can be re-written as
\begin{align}
\nonumber y[n] = \sum_{p=1}^P\sum_{m=0}^M ( &\theta_{p,m}^{(0)}x[n-m]\left|x[n-m]\right|^{p-1} + \\ &\theta_{p,m}^{(1)}s[n]x[n-m]\left|x[n-m]\right|^{p-1} ) \label{model_structure_mp_new}
\end{align}
which results in the parameters $\bm{\theta}$ to be linear with respect to the output. The output can thus be written as
\begin{align}
\mathbf{y} = \mathbf{\left[ \begin{array}{cc} \mathbf{H_x} &  \mathbf{S}\mathbf{H_x}\end{array} \right]}\left[ \begin{array}{cc} \bm{\theta}^{(0) T} &  \bm{\theta}^{(1) T}\end{array} \right]^T
\end{align}
where $\mathbf{S}$ is the diagonal matrix of vector $\mathbf{s}$, and $\mathbf{H_x}$ (for example for an MP with $P=2, M=1$) is
\[ \left[ \begin{array}{cccc}
x[1] & x[1]|x[1]|&0&0\\
x[2] & x[2]|x[2]|&x[1]&x[1]|x[1]|\\
\vdots & \vdots & \vdots & \vdots \\
x[n] &  x[n]|x[n]|&x[n-1] &  x[n-1]|x[n-1]|\end{array} \right]\]

Since now for a known $s[n]$, the model is linear with respect to the parameters $\theta$, the unknown parameters $[\bm{\theta}^{(0) T} \mbox{ } \bm{\theta}^{(1) T}]^T$ can be calculated with a normal least squares technique and written as
\begin{align*}
[\hat{\bm{\theta}^{(0) T}} \mbox{ } \hat{\bm{\theta}^{(1) T}}]^T = \end{align*}
\begin{align} \left(\mathbf{\left[ \begin{array}{cc} \mathbf{H_x} &  \mathbf{S}\mathbf{H_x}\end{array} \right]}^H\mathbf{\left[ \begin{array}{cc} \mathbf{H_x} &  \mathbf{S}\mathbf{H_x}\end{array} \right]}\right)^{-1}
\mathbf{\left[ \begin{array}{cc} \mathbf{H_x} &  \mathbf{S}\mathbf{H_x}\end{array} \right]}^H\mathbf{y}_{\text{meas}}
\label{thetaident}
\end{align}

\subsection{Filter parameters}
In Step 5, since $[\alpha,\beta]$ appear nonlinearly in the output once $\theta$ are fixed, we follow the nonlinear Gauss-Newton method from \cite[p. 260]{kay} for identification. This requires an additional set of iterations, inside the original iteration loop. The iterative steps can be written as
\begin{align}
\nonumber [\alpha_{k+1},\beta_{k+1}] &= [\alpha_{k},\beta_{k}]\\ &+\lambda\left(\left(\mathbf{\Psi}^H\mathbf{\Psi}\right)^{-1}\mathbf{\Psi}^H(\mathbf{y}_{\text{meas}}-\mathbf{y}([\alpha_{k},\beta_{k}]))\right),
\label{a}
\end{align}
where $\lambda$ is the dampening parameter, $\mathbf{y}([\alpha_{k},\beta_{k}])$ is the output of the model with $[\alpha_{k},\beta_{k}]$ as the parameters of the filter, and $\bm{\Psi}$ is the Jacobian matrix defined by
\begin{align}
\mathbf{\Psi} = \left[ \begin{array}{cc}
\frac{\partial{y([\alpha_{k},\beta_{k}])[1]}}{\partial{\alpha}} & \frac{\partial{y([\alpha_{k},\beta_{k}])[1]}}{\partial{\beta}}\\
\frac{\partial{y([\alpha_{k},\beta_{k}])[2]}}{\partial{\alpha}} & \frac{\partial{y([\alpha_{k},\beta_{k}])[2]}}{\partial{\beta}}\\
\vdots & \vdots \\
\frac{\partial{y([\alpha_{k},\beta_{k}])[n]}}{\partial{\alpha}} & \frac{\partial{y([\alpha_{k},\beta_{k}])[n]}}{\partial{\beta}}\end{array} \right]
\end{align}
where,
\begin{align}
\nonumber &= \frac{\partial \mathbf{y}([\alpha_{k},\beta_{k}])[n]}{\partial \mathbf{\alpha}} \\
& =\left(s[n-1] + \alpha\frac{\partial \mathbf{y}([\alpha_{k},\beta_{k}])[n-1]}{\partial \mathbf{\alpha}}\right)\theta^{(1)}
\end{align}
and
\begin{align}
\nonumber &= \frac{\partial \mathbf{y}([\alpha_{k},\beta_{k}])[n]}{\partial \mathbf{\beta}}  \\
&=\left(|x[n-1]|^2 + \alpha\frac{\partial \mathbf{y}([\alpha_{k},\beta_{k}])[n-1]}{\partial \mathbf{\beta}}\right)\theta^{(1)}
\end{align}
It should be noticed that the derivatives are computed recursively, i.e., to compute the derivative at sample $n$ we need the derivative of sample $n-1$.

The iterations are continued until the parameters converge, and then we us these parameters in Step 3 to update the estimate for $s[n]$, until to overall performance converges. This process will be shown with an example in the Section IV.

\section{Results and Analysis}
In this section, we present identification, the behavioral modeling performance, and digital predistortion using the proposed behavioral modeling technique.

\subsection{Measurement setup}
In order to test the validity of the modeling formulation, the following measurement setup is used. The modulator used
is an Agilent E4438C vector signal generator (VSG) and the data is captured with an Agilent N9030A PXA signal analyzer. The baseband I/Q data is generated in a computer and
downloaded to VSG. The VSG modulates the data to an RF carrier and
in order to have enough input power for the PA under test, fed
through a preamplifier. This signal is then fed to a $100$ W LDMOS Doherty power amplifier from NXP semiconductors and the output is captured by the PXA with a sampling rate of 30.72 MSamples/s. The
PXA sends the down-converted baseband I/Q data to the PC and Matlab. All devices are connected by
GPIB and triggered in synch.

\subsection{Test signal}
\begin{figure}
\centering
\includegraphics[width=0.85\columnwidth]{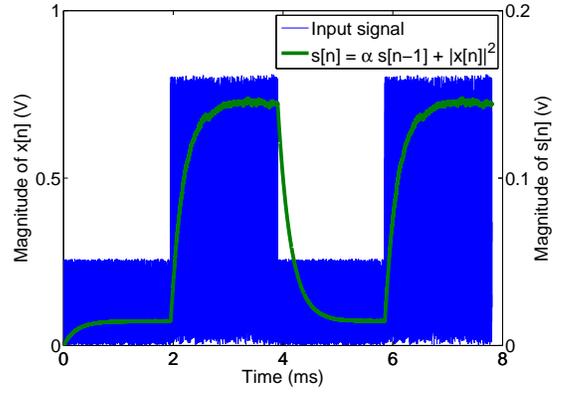}
\caption{The amplitude of the bursty data used to evaluate the long term memory performance of the proposed model. The magnitude of the PA long term memory state is also shown for comparison, as calculated from Section IV. C. } \label{input}
\end{figure}

In order to mimic bursty usage patterns in future generation systems, settings from the 3rd Generation Partnership Project (3GPP) standard for Long Term Evolution (LTE) \cite{lte} are used to construct an input signal. 4 sub-frames (with 2ms for each) are constructed with a 10 dB power change, which represents a sudden increase in input power to the PA, e.g, when the user is downloading a webpage or when a dynamic power allocation algorithm changes the average power of the transmitter. Fig.~\ref{input} shows the amplitude of the 4 MHz white Gaussian noise test signal.

The long term memory state of the PA (calculated for Section IV.C) is also shown. By observing this state variable, it is expected that the PA behavior is affected during a significant time after the transition between the high/low power regions, which cannot be modeled using common behavioral modeling techniques.

\subsection{Performance metrics}
For communication signals similar to Fig.~\ref{input}, it is important to note that the normalized mean squared error (NMSE) is not well suited to evaluate model performance. This is because during the identification stage, if the entire data set is used, the model parameters will be biased towards towards the high power segments. Additionally, using the NMSE will result in averaging the model over the high and low power segments, and will neglect to evaluate the performance during the sudden transition in power.

In order to address this issue, in this work, additional experiments are also conducted to evaluate the instantaneous NMSE performance performance. First the parameters of the model are identified using the entire data set (containing both high and low power parts), and then the mean squared error (MSE) between the model and the measurements is evaluated on a validation set in blocks of 4000 samples (which correspond to blocks of around 0.13 ms), and normalized by the variance of the entire input signal to construct the instantaneous NMSE. The maximum value of this instantaneous NMSE is reported as the maximum NMSE (which corresponds to the worst-case NMSE).

\subsection{Identification}
\begin{figure}
\centering
\includegraphics[width=0.85\columnwidth]{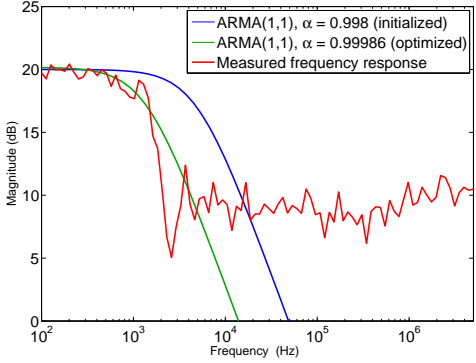}
\caption{Estimated long term memory frequency response of the low--pass filter in Fig.\ref{setup1}, assuming a linearized power amplifier model. The estimated initial and optimized ARMA(1,1) models are also shown.} \label{spectrum}
\end{figure}

After initializing the ARMA parameters using the technique presented in the Appendix, the resulting ARMA(1,1)
filter frequency response used for initialization is shown in Fig.~\ref{spectrum}. It should be noted that the 
magnitude results for frequencies $>5$kHz in Fig.~\ref{spectrum} are limited by measurement noise, which is 
quite high for this kind of measurements. The behavior of the long term memory frequency response is, however, 
expected to present a clear low pass behavior also beyond the cutoff frequency. In the case of the FIR, a full
search of possible values for $N$ is needed.

\begin{figure}
\centering
\includegraphics[width=0.85\columnwidth]{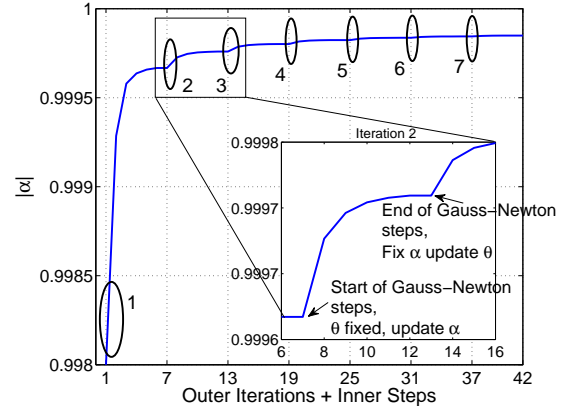}
\caption{The magnitude of the pole of the ARMA(1,1) model over iterations. The numbers in the figure show the outer iteration number. $\theta$ are fixed in the inner steps, and $\alpha$ is fixed in the iteration numbers denoted by the circles. The base model is an MP with $P = 5$ and $M = 2$. The dampening parameter $\lambda$ is equal to 0.4. The inlay shows the convergence between during the second iteration when $\bm{\theta}$ is fixed.} \label{alpha}
\end{figure}

Fig.~\ref{alpha} shows the change in the magnitude of the pole of an ARMA(1,1) model during the iterations. From the figure it can be noticed that once $\bm{\theta}$ is fixed, $\alpha$ from (\ref{a}) converges after 4-6 Gauss-Newton steps. Once $[\alpha,\beta]$ converge, they are fixed and used to compute a new set of $\bm{\theta}$. It can be noted that the entire algorithm converges after 5 iterations between $[\alpha,\beta]$ and $\theta$. It can also be noticed that the biggest change in the amplitude of the pole is after $\theta$ are fixed and the Gauss-Newton steps begin.

Effectively, using $|\alpha| = 0.99986$ (the converged value of the pole) corresponds to a memory depth of approximately 8000 samples (0.25 ms), which can not be modeled using normal MP, GMP or similar techniques due to the explosion in the number of parameters. It was also noticed that, although the algorithm allowed the pole to be complex--valued, the optimization procedure resulted in a real--valued pole which means the long term memory to phase component is negligible. The resulting spectrum of the optimized ARMA(1,1) is shown in Fig.~\ref{spectrum}, which shows a slightly lower cut-off frequency for the long term memory effect than the initial filter.

\subsection{Behavioral modeling performance}
In this section, the behavioral modeling performance of the proposed modeling technique is evaluated, using both the average NMSE and the worst-case NMSE.

The modeling performance from only using the initialization from Section IV.C for the proposed model with an MP(5,2) as its basis model structure -- denoted as LT-MP(5,2) + ARMA(1,1) -- is -49.0 dB NMSE. Compared to an MP(5,4) model with the same number of parameters, which has -48.1 dB NMSE, the proposed technique shows around 0.9 dB modeling performance improvement.

\begin{figure}
\centering
\includegraphics[width=0.85\columnwidth]{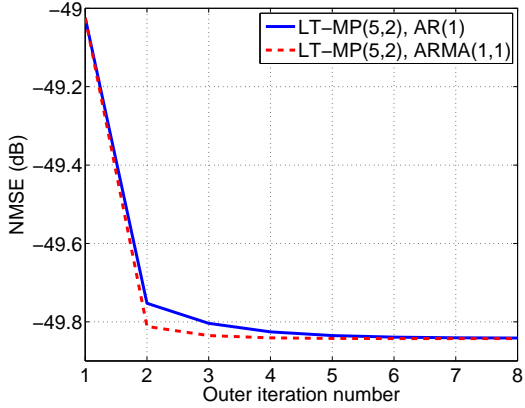}
\caption{NMSE vs the outer iteration number for two long term memory filters $|G(\omega)|$, AR(1) and ARMA(1,1).} \label{NMSE_converge}
\end{figure}

Fig.~\ref{NMSE_converge} shows the NMSE of two models using an AR(1) or ARMA(1,1) filter for the long term memory state, during the iterative identification algorithm. It can be noticed that the NMSE shows a further 0.8 dB performance improvement from the initial filter calculated from the two tone measurement. The average NMSE is 0.3 dB better than the FIR filter presented in \cite{soltani2012} and around 1.8 dB better than the MP(5,4) model that has the same number of parameters. It can be also noticed that using an ARMA(1,1) results in a slightly improved convergence compared to using the simpler AR(1) model.

In Table \ref{tableBH} the model performance is compared with different models and different parameter combinations. In the case of the Volterra model, it was noticed that in open tests increasing the memory length did not improve the modeling performance, and hence those results were not included. It can be noticed that the NMSE performance of regular MP, GMP and Volterra models can be improved by around 1.8 -- 2 dB when applied with the proposed long-term modeling scheme. The performance using an AR filter instead of an FIR filter shows around 0.2 dB improvement for the different models. The average out of band modeling performance is also shown with the adjacent channel error power ration (ACEPR) \cite{soltani}, and the performance is improved by around 2.5-3 dB.

\begin{figure}
\centering
\includegraphics[width=0.85\columnwidth]{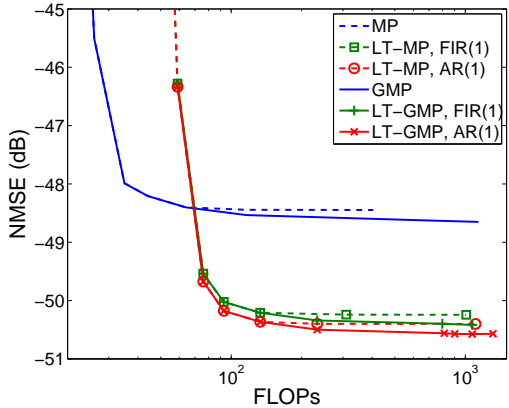}
\caption{The accuracy/complexity tradeoff for different models.} \label{flops}
\end{figure}

Fig.~\ref{flops} shows the accuracy/complexity tradeoff for the different models as proposed in \cite{soltani}. It can noticed that except for the low floating point (FLOP) region, the proposed modeling technique improves the performance by around 2 dB for both the model with an MP basis and the model with a GMP basis.

As mentioned previously, utilizing solely the average NMSE for model evaluation is not sufficient to compare the model performance. In order to analyze the instantaneous performance, the maximum NMSE is also given in Table \ref{tableBH}. It can be noticed that in the case of MP and GMP, the worst-case NMSE falls to -43.2 dB and in the case for Volterra to -36.9. Utilizing the proposed technique, the worst-case NMSE is improved by around 4.5--5 dB.
\begin{table}
\caption{Comparison of behavioral modeling performance for the proposed modeling technique and MP and GMP. Model orders are shown in the parenthesis and are chosen for relatively similar complexity.} \centering  {\label{tableBH}
\begin{tabular}{ l || c | c | c | c}
  Model& NMSE & Max NMSE & ACEPR &  Number of\\
  & (dB) & (dB)& (dB) & parameters\\
  \hline
  \hline
  \hline
  MP(7,10)& -48.2 & -43.2 & -61.5 & 77\\
  GMP(7,3,1)& -48.5 & -43.2 & -61.7& 123\\
  Volterra (7,1) & -45.3 & -36.9 & -57.7 & 40\\
  \hline
  \hline
  FIR(1) & & &\\
  \hline
  LT-MP(7,4)& -50.0 & -47.8& -64.5 & 78\\
  LT-GMP(7,2,1)&-50.2 & -47.7& -64.7 & 118\\
  LT-Volterra (7,1)& -46.5 & -41.3 & -60.0 & 81\\
  \hline
  \hline
  AR(1)&&&\\
  \hline
  LT-MP(7,4) & -50.2 & -48.2 & -64.8& 78\\
  LT-GMP(7,2,1) & -50.4 & -48.0 & -65.0& 118\\
  LT-Volterra (7,1) & -47.6 & -41.5 & -60.2 & 81\\
  \hline
\end{tabular}}
\end{table}

%
%

\begin{figure}
\centering
\includegraphics[width=0.85\columnwidth]{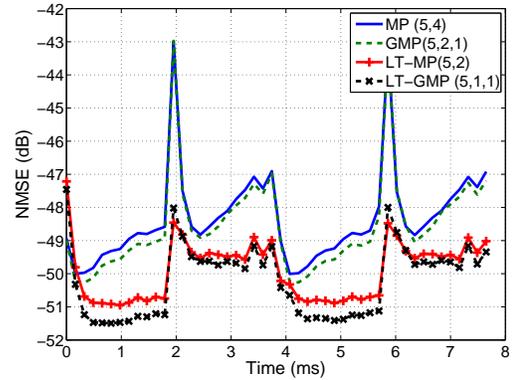}
\caption{NMSE of modeling performance vs time. Model orders are in parenthesis and are chosen for relatively similar complexity of each model.} \label{NMSE_performance}
\end{figure}

Fig.~\ref{NMSE_performance} shows the instantaneous NMSE over time. The proposed model shows a consistent 2-3 dB better modeling performance than conventional models in both high and low power segments. The instantaneous performance improvement can especially be noticed in transitions where the proposed model is able to track the sudden change in PA characteristics better.

It is also interesting to note that the performance of the proposed model for the beginning of the cycle is slightly worse (in calculating the max NMSE this section is ignored so that the outcome does not depend on the initial state). This is because the long term memory estimate is initialized at zero, and takes some time to ramp up to the correct values. This effect is not noticed in the second cycle as the long term memory estimate is consistent now. In calculating the maximum NMSE this initialization effect is ignored.

\subsection{Digital Pre-Distortion Linearization}
In this section, the behavioral modeling technique is used for digital predistortion. For these comparisons, special care was taken to maintain a constant average output power for both PA without and the PA with predistortion as specified in \cite{zhudpd}. For DPD, the parameters of $G(\omega)$ are identified from the behavioral model using the iterative technique mentioned previously, and then kept fixed. Once these parameters are fixed, the DPD can be done in the normal fashion, by reversing the input and output, identifying an inverse model of the PA and using it to predistort the data. The results of DPD are shown after a few iterations when the performance converges.

For comparison, DPD results of using an MP + AR filter and a GMP + AR filter are shown in Table \ref{tableDPD}.
\begin{table}
\caption{Comparison of DPD for the proposed modeling technique and MP and GMP. Model orders (in parenthesis) are chosen for relative similar complexity.} \centering  {\label{tableDPD}
\begin{tabular}{ l || c | c | c }
  Model& NMSE (dB) & ACPR (dB) & Max NMSE (dB)\\
  \hline
  \hline
  MP(7,8)& -47.8 & -49.6&-43.6\\
  GMP(7,4,1)& -49.2 & -51.8&-45.9\\
  LT-MP(7,4), AR(1)& -49.9& -54.5&-47.4\\
  LT-GMP(7,2,1), AR(1)&-49.9&-53.2&-47.3\\
  \hline
\end{tabular}}
\end{table}
It can be noticed that using the proposed method improves the average NMSE performance of MP by around 2 dB and around 0.7 dB for the GMP model. The out-of-band ACPR is also improved by around 2 dB for the GMP model, and almost 6 dB for the MP. The maximum NMSE is also improved using the proposed model by around 2-4 dB, which can be critical in most applications.

\begin{figure}
\centering
\includegraphics[width=0.85\columnwidth]{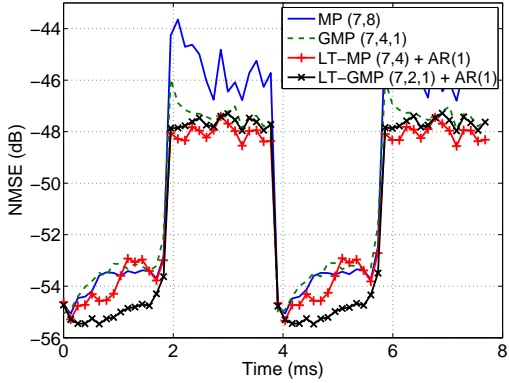}
\caption{Instantaneous NMSE after predistortion for the different models.} \label{dpd}
\end{figure}

Fig.~\ref{dpd} shows the instantaneous NMSE computed over blocks of 4000 samples after  predistortion for the different models above. It can be noticed that the LT-MP successfully improves the linearity of the system by around 3-4 dB in the high power segment and the LT-GMP model improves the NMSE by approximately 1-2 dB.

\begin{figure}
\centering
\includegraphics[width=0.85\columnwidth]{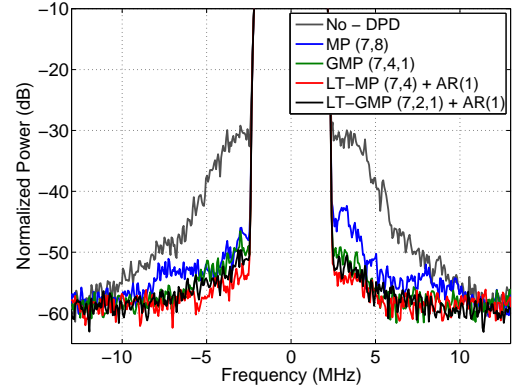}
\caption{Spectrum of the output of the PA with predistortion for different models at the switching instance.} \label{spectrum_dpd}
\end{figure}

Fig.~\ref{spectrum_dpd} shows the spectrum of the PA output for different models in the switching transition block. It can be noticed that the proposed modeling technique improves the worst--case (during power change) adjacent channel power ratio (ACPR) of the MP model by approximately 6-10 dB. In a traditional DPD linearization architecture, parameter adaptation would be necessary in order to avoid the degradation in performance. Utilizing the proposed technique can help lessen the burden on the traditional adaptation hardware.

\begin{figure}
\centering
\includegraphics[width=0.85\columnwidth]{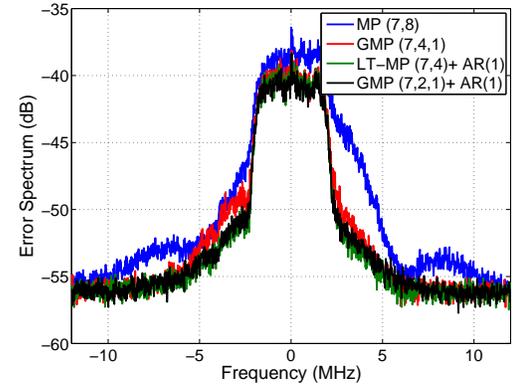}
\caption{Error spectrum of the output of the PA with predistortion for different models at the switching instance.} \label{error_spectrum}
\end{figure}

The error spectrum is shown in Fig.~\ref{error_spectrum}, to further analyze the out of band performance. It can be noticed that the LT-MP and LT-GMP models outperform the normal counterparts for a relatively similar complexity.

\begin{figure}
\centering
\includegraphics[width=0.85\columnwidth]{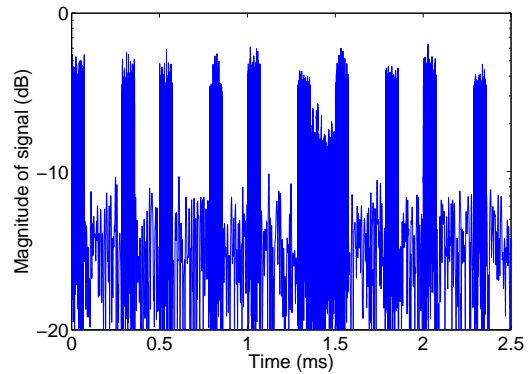}
\caption{Time record of the LTE data signal used.} \label{ltedata}
\end{figure}

In the next experiment, the proposed technique is compared to traditional techniques with a 20 MHz 
Long Term Evolution (LTE) test data E-UTRA Testmodel TM2 \cite{3gpp}. A time record of the signal is shown in
Fig.\ref{ltedata}. This signal is used to test the dynamic range of base stations, the error vector magnitude 
and the frequency error.

\begin{figure}
\centering
\includegraphics[width=0.85\columnwidth]{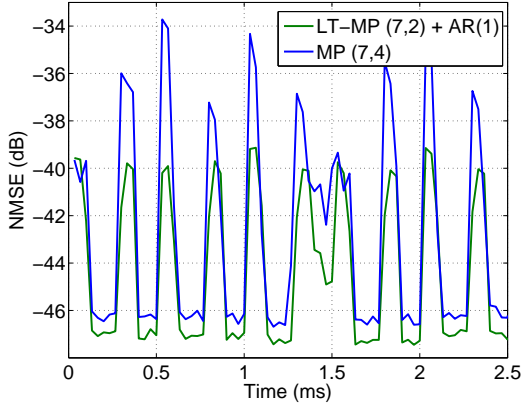}
\caption{Instantaneous NMSE after predistortion for the different models.} \label{ltedpd}
\end{figure}

The resulting instantaneous NMSE DPD performance for a traditional MP(7,4) and the LT-MP(7,2) + AR filter for 
this data is shown in Fig.\ref{ltedpd}. It can be noticed that the proposed model improves the overall performance
for this data by 4-5 dB for the relative same computational complexity.

\subsection{Model Generalization}
\begin{figure}
\centering
\psfrag{x}[c][c][1]{$x[n]$}\psfrag{y}[c][c][1]{$y[n]$}\psfrag{s}[c][c][1]{$s_1[n]$}\psfrag{A}[c][c][1]{$\mathbf{H_x }\bm{\theta}\mathbf{(s_1,s_2,\cdots,s_n)}$}
\psfrag{C}[c][c][1]{$G_1\left(\omega\right)$}\psfrag{B}[c][c][1]{$|\cdot|^2$}
\psfrag{t}[c][c][1]{$s_2[n]$}\psfrag{D}[c][c][1]{$G_2\left(\omega\right)$}
\psfrag{v}[c][c][1]{$s_n[n]$}\psfrag{E}[c][c][1]{$G_n\left(\omega\right)$}
\includegraphics[width=0.8\columnwidth]{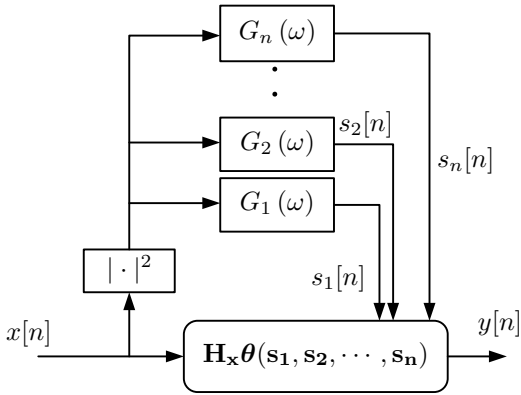}
\caption{Tracking multiple states in the PA by extending the proposed model.} \label{setup2}
\end{figure}

The model can easily be extended to track more than one state in the PA. These states can represent different time constants in the PA architecture design. These states can also be filtered ARMA, AR or FIR models, as shown in Fig.~\ref{setup2}. In this work, as an example, three AR(1) filters are used to track different states in the PA. The first state is initialized from the same two tone measurement setup as explained previously. Since the maximum memory length was captured by the two tone measurements, the two other states are initialized for shorter memory lengths. The same iterative algorithm as before can be used, iterating between keeping $\bm{\theta}$ fixed and identifying $[\alpha,\beta]_1$, $[\alpha,\beta]_2$, $[\alpha,\beta]_3$ one at a time (in a greedy fashion), and keeping the three states fixed and identifying $\bm{\theta}$. 

\begin{figure}
\centering
\includegraphics[width=0.85\columnwidth]{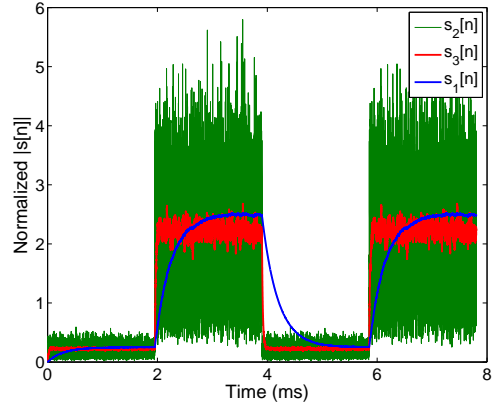}
\caption{The three long term memory states $\bm{s}$ computed from the iterative algorithm.} \label{s3}
\end{figure}

It was noticed that after convergence, while the first filter had a time constant of around 8000 samples, the second filter modeled 20 ($\alpha_2 = 0.949$) and the third filter modeled around 110 samples ($\alpha_3 = 0.991$). These three different states are shown in Fig.~\ref{s3}. Notice that $s_2[n]$, which is represented by the AR(1) filter with $\alpha_2$, has more instantaneous changes in the power while $s_1[n]$ with the AR(1) filter with $\alpha_1$ has a much slower changing power profile change.

\begin{figure}
\centering
\includegraphics[width=0.85\columnwidth]{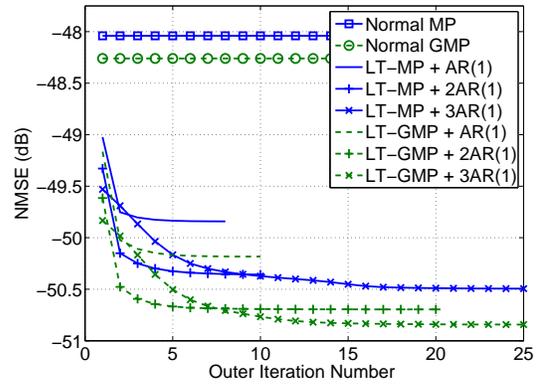}
\caption{NMSE vs the outer iteration number for different modeling paradigms. The normal MP and GMP are also shown for comparison. Model orders are chosen to keep relatively similar complexity.} \label{3IIR}
\end{figure}

In order to compare using one, two and three states' model performance, the following experiment is conducted. An MP and GMP model are combined with one AR(1), two AR(1) and three AR(1) filters to improve modeling accuracy. The result of the iterative identification algorithm is shown in Fig.~\ref{3IIR}. It can be noticed that using two AR filters instead of one results in an additional 0.5 dB improvement for both the MP and GMP models. Further using an additional filter improves the performance by around 0.2 dB. It can be noticed that compared to the original MP and GMP models with comparable complexities, the performance is improved by around 1.5 dB for one AR filter, 2.3 dB for two AR filters and 2.5 dB for three AR filters for both the MP and GMP model. Using more than one state increases the amount of iterations needed for convergence and the initialization sensitivity compared to the single state case, but the convergence is still relatively fast.

\section{Conclusions}
In this paper, a black--box modeling technique for modeling long term memory effects is presented. Identification of these effects using an iterative algorithm is presented and it is shown that the model converges after a few iterations. We have demonstrated that the model is well suited to handle bursty data which is becoming more and more common in wireless systems.

The results show that by linearizing the parameters with respect to a long term memory state, it is possible to accurately track abrupt changes in PA characteristics and improve the average NMSE by around 1.5--2.5 dB and the worst-case NMSE by around 4-5 dB for commonly used behavioral models in the literature. The model is used for digital predistortion and the results show 1-2 dB average in-band and 4-8 dB worst-case out of band performance improvement.

\renewcommand{\theequation}{A.\arabic{equation}}
\setcounter{equation}{0}

\section*{Appendix}\label{append}
The procedure to characterize the long term memory effect for initialization is presented in this Appendix. In order to estimate the frequency response $G(\omega)$, from the two tone measurements, a simple first order model is constructed assuming no short term memory component.
\begin{align}
y[n,\omega_0] = \theta_0^{(0)}x[n,\omega_0] + \theta_0^{(1)}s[n,\omega_0]x[n,\omega_0] \label{math1append}
\end{align}
Since a single offset frequency is considered the filter response at that frequency can be obtained and $s[n,\omega_0]$ can be written as
\begin{align}
s[n,\omega_0] = A^2 + B^2 + AB|G(\omega_0)|\Phi(\omega_0)\left(e^{-j\omega_0n}+e^{j\omega_0n}\right)
\end{align}
where $|G(\omega)|$ is the amplitude of the frequency response of the filter, and $\Phi(\omega_0) = e^{-j\phi(\omega_0)}$ is the phase. We can assume that the filter gain is normalized at zero.
Also, for simplicity, it can be assumed that the output of the filter is real-valued, i.e., $G(\omega) = G(-\omega)^*$ and $\Phi(\omega_0)=1$.

Given this setup, $y[n]$ can be calculated as
\begin{eqnarray}
\nonumber y[n,\omega_0] = \theta_0^{(0)}x[n,\omega_0] + \theta_0^{(1)}s[n,\omega_0]x[n,\omega_0]\\
\nonumber = \theta_0^{(0)}\left(Ae^{-j\omega_0n} + B\right) + \theta_0^{(1)}(e^{-j2\omega_0n}\left[A^2B|G(\omega_0)|\right] + \\
\nonumber e^{-j\omega_0n}\left[A^3+AB^2+AB^2|G(\omega_0)|\right]\\
\nonumber + A^2B + B^3 + A^2B|G(\omega_0)| + e^{j\omega_0n}\left[AB^2|G(\omega_0)|\right])
\end{eqnarray}
Rewriting the terms based on frequency components
\begin{eqnarray}
\nonumber y[n,\omega_0] = \\
\nonumber e^{-j2\omega_0n}\left[A^2B|G(\omega_0)|\theta_0^{(1)}\right] \\
\nonumber + e^{-j\omega_0n}\left[A\theta_0^{(0)}+ (A^3+AB^2+AB^2|G(\omega_0)|)\theta_0^{(1)}\right]\\
+ \left[B\theta_0^{(0)} + (A^2B + B^3 + A^2B|G(\omega_0)|)\theta_0^{(1)}\right] \\
\nonumber + e^{j\omega_0n}\left[AB^2|G(\omega_0)|\theta_0^{(1)}\right] \label{redmath1append}
\end{eqnarray}

The magnitude of these frequency components can be determined from simple spectrum measurements. The unknown parameters of $\theta_0^{(0)}$, $\theta_0^{(1)}$ and $|G(\omega_0)|$ and can be identified by substitution in these tones. By using the three tones of $-\omega_0$, $0$ and $+\omega_0$ at each frequency offset, we can construct the following matrix for least-squares identification of $\theta_0^{(0)}$ and $\theta_0^{(1)}$,
\begin{align}
\nonumber \begin{pmatrix}
  |Y[0]|_0 \\
  |Y[\omega_1]|_0-|Y[\omega_1]|_{\omega_1}\\
  \vdots\\
  |Y[\omega_{\text{max}}]|_0-|Y[\omega_{\text{max}}]|_{\omega_{\text{max}}}
 \end{pmatrix} =\\
 \begin{pmatrix}
  (A+B)&(A+B)^3 \\
  B&B^3+A^2B\\
  \vdots&\vdots\\
  B&B^3+A^2B
 \end{pmatrix}
 \begin{pmatrix}
  \theta_0^{(0)} \\
  \theta_0^{(1)}
 \end{pmatrix}
\label{thetamatrixappend}
\end{align}
where each row in the matrix is one measurement with frequency $\omega_k$, $|Y(\omega_k)|_0$ is the measured amplitude of the spectrum at the frequency zero frequency offset and $|Y(\omega_k)|_{\omega_k}$ at the $\omega_k$ frequency offset. The equation for constructing $|G(\omega)|$ can be written as
\begin{align}
|G(\omega)| = \frac{|Y[n,\omega]|_0-\left(B^3+A^2B\right)\theta_0^{(1)}-B\theta_0^{(0)}}{A^2B\theta_0^{(1)}}
\label{g2eqappend}
\end{align}

Once the frequency response is obtained from the two tone tests, traditional frequency domain estimation techniques 
like \cite[p. 230]{ljung} for identifying the linear filter in the long--term memory path (for example the ARMA 
filter) can be employed. It can be noted that the zeros and poles of the AR and ARMA models can be forced to 
be real-valued to maintain a real--valued state parameter, which could represent tracking changes in the 
average power in the input signal. In this work we have relaxd this condition and let zeros and poles to be 
complex valued as well, which may result in an input signal power to phase component in the long term memory.

\section*{Acknowledgement}
This research has been carried out in GigaHertz Centre in a joint project financed by the Swedish Governmental Agency
for Innovation Systems (VINNOVA), Chalmers University of Technology, Ericsson AB, Infineon Technologies Austria AG, NXP Semiconductors BV, and Saab AB.

\ifCLASSOPTIONcaptionsoff
  \newpage
\fi

\bibliographystyle{IEEEtran}
\bibliography{main}

\begin{thebibliography}{10}
\providecommand{\url}[1]{#1}
\csname url@samestyle\endcsname
\providecommand{\newblock}{\relax}
\providecommand{\bibinfo}[2]{#2}
\providecommand{\BIBentrySTDinterwordspacing}{\spaceskip=0pt\relax}
\providecommand{\BIBentryALTinterwordstretchfactor}{4}
\providecommand{\BIBentryALTinterwordspacing}{\spaceskip=\fontdimen2\font plus
\BIBentryALTinterwordstretchfactor\fontdimen3\font minus
  \fontdimen4\font\relax}
\providecommand{\BIBforeignlanguage}[2]{{%
\expandafter\ifx\csname l@#1\endcsname\relax
\typeout{** WARNING: IEEEtran.bst: No hyphenation pattern has been}%
\typeout{** loaded for the language `#1'. Using the pattern for}%
\typeout{** the default language instead.}%
\else
\language=\csname l@#1\endcsname
\fi
#2}}
\providecommand{\BIBdecl}{\relax}
\BIBdecl

\bibitem{pedro}
J.~C. Pedro and S.~A. Maas, ``A comparative overview of microwave and wireless
  power-amplifier behavioral modeling approaches,'' \emph{IEEE Trans. Microw.
  Theory Tech.}, vol.~53, no.~4, pp. 1150--1163, 2005.

\bibitem{soltani}
A.~S.Tehrani, H.~Cao, S.~Afsardoost, T.~Eriksson, M.~Isaksson, and C.~Fager,
  ``A comparative analysis of the complexity/accuracy tradeoff in power
  amplifier behavioral models,'' \emph{IEEE Trans. Mircow. Theory Tech.},
  vol.~58, pp. 1510--1520, 2010.

\bibitem{isaksson}
M.~Isaksson, D.~Wisell, and D.~Ronnow, ``A comparative analysis of behavioral
  models for {RF} power amplifiers,'' \emph{IEEE Trans. Microw. Theory Tech.},
  vol.~54, no.~1, pp. 348--359, 2006.

\bibitem{vuolevi}
J.~H.~K. Vuolevi, T.~Rahkonen, and M.~J.~P. A., ``Measurement technique for
  characterizing memory effecs in {RF} power amplifiers,'' \emph{IEEE Trans.
  Mircowave Theory Tech.}, vol.~49, pp. 1381--1389, Aug. 2001.

\bibitem{ngoya}
E.~Ngoya, C.~Quindroit, and J.~Nebus, ``On the continuous-time model for
  nonlinear-memory modeling of rf power amplifiers,'' \emph{IEEE Trans. Microw.
  Theory Tech.}, vol.~57, no.~12, pp. 3278 --3292, dec. 2009.

\bibitem{ku2003}
H.~Ku and K.~J. S., ``Behavioral modeling of nonlinear {RF} power amplifiers
  considering memory effects,'' \emph{IEEE Trans. Mircow. Theory Tech.},
  vol.~51, pp. 2495--2504, Dec. 2003.

\bibitem{thermal_boumaiza}
S.~Boumaiza and G.~F. M, ``Thermal memory effects modeling and compensation in
  {RF} power amplifiers and predistortion linearizers,'' \emph{IEEE Trans.
  Mircowave Theory Tech.}, vol.~51, pp. 2427--2433, 2003.

\bibitem{bosch}
W.~Bosch and G.~Gatti, ``Measurement and simulation of memory effects in
  predistortion linearizers,'' \emph{IEEE Trans. Mircow. Theory Tech.},
  vol.~37, no.~12, pp. 1885 --1890, dec 1989.

\bibitem{coda}
\BIBentryALTinterwordspacing
S.~Smith. (2010, March) Mobile internet handsets in the us: A comprehensive
  assessment, with forecasts to 2015. [Online]. Available:
  \url{http://www.codaresearch.co.uk/usmobileinternet/index.htm}
\BIBentrySTDinterwordspacing

\bibitem{thermal_mazeau}
J.~Mazeau, R.~Sommet, D.~Caban-Chastas, E.~Gatard, R.~Quere, and Y.~Mancuso,
  ``Behavioral thermal modeling for microwave power amplifier design,''
  \emph{IEEE Trans. Mircowave Theory Tech.}, vol.~55, pp. 2290--2297, Nov.
  2007.

\bibitem{crespo2010}
C.~Crespo-Cadenas, J.~Reina-Tosina, and M.~Madero-Ayora, ``Study of a power
  amplifier behavioral model with nonlinear thermal effects,'' in
  \emph{Microwave Conference (EuMC), 2010 European}, sept. 2010, pp. 1138
  --1141.

\bibitem{crespo2011}
C.~Crespo-Cadenas, J.~Reina-Tosina, and M.~J. Madero-Ayora, ``An equivalent
  circuit-based approach to behavioral modeling of long-term memory effects in
  wideband amplifiers,'' \emph{Microw. Opt. Techn. Letters}, vol.~53, no.~10,
  pp. 2278--2281, 2011.

\bibitem{soltani2012}
A.~S.Tehrani, T.~Eriksson, and C.~Fager, ``Modeling of long term memory effects
  in rf power amplifiers with dynamic parameters,'' in \emph{To appear Proc.
  IEEE MTT-S Int. Micr. Symp. Dig.}, 2012.

\bibitem{kim}
J.~Kim and K.~Konstantinou, ``Digital predistortion of wideband signals based
  on power amplifier model with memory,'' \emph{Electron. Lett.}, vol.~27, pp.
  1417--1418, 2001.

\bibitem{morgan}
D.~Morgan, Z.~Ma, J.~Kim, M.~Zierdt, and J.~Pastalan, ``A generalized memory
  polynomial model for digital predistortion of rf power amplifiers,''
  \emph{IEEE Trans. Sig. Proc.}, vol.~54, pp. 3852 --3860, oct. 2006.

\bibitem{ngoya2009}
E.~Ngoya, C.~Quindroit, and J.~Nebus, ``Improvements on long term memory
  modeling in power amplifiers,'' in \emph{Proc. IEEE MTT-S Int. Micr. Symp.
  Dig.}, 2009, pp. 1357--1360.

\bibitem{kay}
S.~Kay, \emph{Fundamentals of Statistical Signal Processing, Estimation
  Theory}.\hskip 1em plus 0.5em minus 0.4em\relax Prentice Hall, 1993.

\bibitem{lte}
\BIBentryALTinterwordspacing
J.~Zyren. (2007, Aug,) Overview of the 3gpp long term evolution physical layer.
  [Online]. Available:
  \url{http://www.freescale.com/files/wirelesscomm/doc/whitepaper/3GPPEVOLUTIONWP.pdf}
\BIBentrySTDinterwordspacing

\bibitem{zhudpd}
A.~Zhu, P.~Draxler, J.~Yan, T.~Brazil, D.~Kimball, and P.~Asbeck, ``Open-loop
  digital predistorter for {RF} power amplifiers using dynamic deviation
  reduction-based volterra series,'' \emph{IEEE Trans. Microw. Theory Tech.},
  vol.~56, no.~7, pp. 1524 --1534, july 2008.

\bibitem{3gpp}
\BIBentryALTinterwordspacing
ETSI. (2006, Dec.) Lte; evolved universal terrestrial radio access (e-utra);
  base station (bs) conformance testing (3gpp ts 36.141 version 8.3.0 release
  8). [Online]. Available:
  \url{http://www.etsi.org/deliver/etsi_ts/136100_136199/136141/08.03.00_60/ts_136141v080300p.pdf}
\BIBentrySTDinterwordspacing

\bibitem{ljung}
L.~Ljung, \emph{System identification: theory for the user, 2nd ed.}\hskip 1em
  plus 0.5em minus 0.4em\relax Prentice-Hall, Inc., 1999.

\end{thebibliography}

\end{document}